\documentclass{sig-alternate-05-2015}

\usepackage{listings}
\usepackage{booktabs} 
\usepackage{tikz}
\usetikzlibrary{calc}
\usepackage{hyperref}
\usepackage{breakurl}
\usepackage{mdframed}
\usepackage{pgfplots}
\usepackage{pgfplotstable}
\usepackage{graphicx}
\usepackage{subcaption}
\usepackage{caption}
\usetikzlibrary{calc}
\usetikzlibrary{patterns}
\usetikzlibrary{shapes.multipart, arrows}
\usetikzlibrary{decorations.pathreplacing}
\usepackage{multirow}
\usepackage[PseudoCode]{algorithm}
\usepackage[noend]{algpseudocode}
\usepackage{color}
\usepackage{xcolor,colortbl}
\usetikzlibrary{shapes,snakes}
\usepackage{enumitem}
\usepackage{balance}
\usepackage{filecontents}

\makeatletter
\def\@copyrightspace{\relax}
\makeatother

\makeatletter
\DeclareRobustCommand*\cal{\@fontswitch\relax\mathcal}
\makeatother

\definecolor{cornellred}{rgb}{0.7, 0.11, 0.11}
\definecolor{arsenic}{rgb}{0.23, 0.27, 0.29}
\definecolor{auburn}{rgb}{0.43, 0.21, 0.1}
\definecolor{charcoal}{rgb}{0.21, 0.27, 0.31}
\definecolor{deepcarrotorange}{rgb}{0.91, 0.41, 0.17}
\definecolor{eggplant}{rgb}{0.38, 0.25, 0.32}

\begin{document}\sloppy

\def\codename{\textsc{TEEKAP}}

\title{Self-Expiring Data Capsule using Trusted Execution Environment}

\author{{Hung Dang, Ee-Chien Chang}\\
School of Computing, National University of Singapore\\
\{hungdang,changec\}@comp.nus.edu.sg}
\maketitle

\section*{Abstract}

Data privacy is unarguably of extreme importance. Nonetheless, there exist various daunting challenges 
to safeguarding data privacy. These challenges stem from the fact that data owners have little control over their data once it has transgressed their local storage and been managed by third  parties whose trustworthiness is questionable at times. Our work seeks to enhance data privacy by constructing a \textit{self-
expiring data capsule}. Sensitive data is encapsulated into a capsule which is 
associated with an access policy and expiring condition. The former  indicates eligibility of functions that can access 
the data, and the latter dictates when the data should become inaccessible to anyone, including the previously 
eligible functions. Access to the data capsule, as well as its dismantling once the expiring 
condition is met, are governed by a committee of independent and mutually distrusting nodes. The pivotal 
contribution of our work is an integration of hardware primitive, state machine replication 
and threshold secret sharing in the design of the self-expiring data 
encapsulation framework. We implement the proposed framework in a system called \codename. Our empirical 
experiments conducted on a realistic deployment setting with the access control committee spanning across 
four geographical regions reveal that \codename\ can process access requests at scale with sub-second latency.

\section{Introduction}
\label{sec:intro}
Data privacy has always been a great topic of interest, and it is even more so in the wake of recent devastating data breach incidents reported to have far-reaching and serious implications on both individual and institutional levels~\cite{fb_scandal,panama,SingHealth_leak}. 
Protecting data privacy, nonetheless, faces with various daunting challenges, as data owners have little control over their data once it has transgressed their local storage. For instance, in the context of cloud services, personal and sensitive data are typically collected, stored, processed and archived by third parties whose trustworthiness are questionable at times (e.g., Facebook data scandal~\cite{fb_scandal}). In another example, social factors ranging from individuals' carelessness to courts' jurisdictional actions increase the likelihood of private data being disclosed against the will of data owners. 

%
We seek to enhance data privacy by empowering the data owners with more control over their sensitive data. 
We focus on two properties, namely access eligibility, and data expiry. 
The first property demands that only selected functions with well-defined eligibility can access and compute on the sensitive data. 
The access control structure must be able to
attest the eligibility of the requesting functions and the assurance that such functions, once attested, do not deviate from their intended behaviours. 
The second property ties the sensitive data to an expiring condition. 
Once such condition is met, the data is rendered  inaccessible by any party (including the previously eligible 
functions)  without explicit action from the data owners, 
parties who archive that data, or any trusted third party or external service. 
More specifically, it is required that the data becomes unattainable permanently 
once it is expired in an autonomous manner. 
This calls for decentralization and autonomy of the expiry mechanism. 
At the same time, it is also necessary to ensure that the eligible functions accessing the data 
prior to its expiration \textit{do not} leak it to any unintended entity, nor persist a copy of the data on their storage. 
In order words, we aim to construct a self-expiring data capsule whose content (i.e., the sensitive data) is accessible to a selected eligible functions prior to its expiry, and to none after its expiry.

We envision that the data encapsulation mechanism described above would be beneficial to privacy protection of a broad range of applications that involve digital content. 
In fact, various applications have adopted the use of expiring/disappearing data objects, for instances Snapchat\footnote{\url{https://www.snapchat.com/}} and 
Instagram\footnote{\url{https://www.instagram.com/}}. 
In these applications, messages or stories disappear after a predefined time window without any action from the end users. Nonetheless, such disappearance are not autonomous, instead triggered by a centralised service provider whose trustworthiness may be questionable at times. This is not desirable, for the service provider becomes a single-point-of-failure.
Another example is self-destruct chip~\cite{self_destruct_chip}. Such a chip can be used to store sensitive data, and can be triggered remotely to shatter into smithereens such that reconstruction is impossible. 
However, the shattering is not autonomous, but requires explicit invocation. 
We elaborate on different use-cases and the needs of self-expiring data capsule in Section~\ref{subsec:usecase}.

Mechanisms for self-expiring data capsule without explicit invocations have been studied previously~\cite{vanish, safevanish}. In a nutshell, these 
approaches create a data capsule by encrypting the data that needs protecting with a randomly 
generated encryption key, and secret-shares~\cite{shamir_secret_share} the key to random nodes in a 
decentralized Distributed Hash Table (DHT)~\cite{DHT}. The self-expiring property is realized by the fact 
that nodes in DHT constantly churn or internally cleanse themselves. This results in shares of the 
encryption key naturally vanishing overtime, thereby rendering the protected data (which is encrypted) 
inaccessible. These approaches, however, base the expiring condition of the protected data only in form 
of timeouts, and do not generalize to other type of condition (e.g., number of times the data has been 
accessed). Furthermore, they require parties that have legitimate access to the protected data prior to 
its expiry to trust each other in a sense that none of them will intentionally leak or permanently 
preserve the plaintext copy of the encapsulated data via out-of-band means. Designing a self-expiring 
data capsule that does not observe the aforementioned limitations remains technically challenging.

This paper presents \codename\ - a framework for self-expiring data capsule that sidesteps the 
limitations described earlier. \codename\ supports generic and flexible user-defined 
access policy and expiring conditions, enforcing data expiry even against parties 
that are eligible to access  the data prior to its expiry. 
The pivotal contribution of our work is an integration of hardware primitives, state machine 
replication and secret-sharing protocol. 
Intuitively, \codename\ comprises a committee of \textit{independent} and \textit{mutually distrusting} nodes that leverage a consensus  protocol~\cite{raft} and secret sharing scheme~\cite{shamir_secret_share} to collectively enforce the access control and expiry criteria dictated by data owners. This eliminates the single-point-of-failure issue.

In order to encapsulate her sensitive data with regard to a certain expiring condition, the data owner first 
encrypts her data using a randomly generated encryption key. She then divides this encryption key into 
shares (e.g., using Shamir secret-sharing~\cite{shamir_secret_share}). Subsequently, she deposits the 
shares and the access policy (e.g., which   function can access the data, together with the expiring 
condition) to nodes in the access control committee. To ensure the security of \codename\, it is crucial 
that committee members preserve the confidentiality of the shares assigned to them, as well as 
consistency of the availability status of the encapsulated data with regard to the expiring condition. For 
instance, if the expiring condition is based on the number of times the data has been accessed, the 
committee members should agree on the access history, thereby agreeing on the availability status of the 
data. To establish such an agreement, committee members in \codename\ relies on consensus protocol to 
ensure their view of the data's availability status converge.

Any party that wishes to access the encapsulated data must first provision a trusted execution 
environment (TEE)~\cite{enclave_formalization} and attest the correct instantiation of the TEE to the access control committee~\cite{sgx_remote_attest}. The committee members then collectively verify the 
eligibility of the request, as well as the availability of the protected data with regard to the 
predefined expiring condition. 
If the expiring condition has been met, the nodes discard the key shares assigned to them, and rejects the 
request. On the other hand, if the access can be granted, the committee members pool their key 
shares to reconstruct the encryption key. It is crucial that the reconstruction of 
the encryption key must be performed securely so that the confidentiality of the key shares are 
preserved, and none of the nodes in the access control committee can obtain the encryption key intact. 
Once constructed, the key is delivered to the TEE of the requesting party. 
The protected data is decrypted and processed exclusively inside the requester's TEE.
The TEE guarantees that the encryption key is discarded there 
after, and no copy of the encryption key nor the plaintext version of the protected data is persisted 
outside of the TEE.

We implement a prototype of our proposed framework using 
Intel SGX processors~\cite{sgx} and SGX SDK~\cite{sgx_sdk}, and empirically evaluate its 
performance in a realistic deployment settings. Our experiments show that an access control committee 
spanning over four separate geographical regions in \codename\ can process access request at scale in 
real-time, incurring sub-second latency. 

In summary, this work makes the following contributions:
\begin{itemize}
\item We describe principle specifications and requirements for self-expiring data encapsulation that supports generic and flexible user-defined expiring conditions.

\item We elaborate on the technical challenges of designing the self-expiring data encapsulation by surveying candidate approaches and pointing out their limitations.

\item We present our framework called \codename\, that addresses the technical challenges and attains the desired requirements. The pivotal contribution underlying \codename\ design is a unique integration of hardware primitives, state machine replication and cryptographic protocol.

\item We implement a prototype of \codename, conduct experiments in realistic deployment setting and empirically show the efficiency of our system. 

\end{itemize}

\section{Preliminaries}
\label{sec:background}

In this section, we give brief overview of the building blocks that we employ in the designing of \codename. Section~\ref{subsec:SGX} describes Intel SGX which is a hardware primitive we use to provision TEE. Section~\ref{subsec:consensus} characterizes key essence of consensus protocols, while Section~\ref{subsec:secre_sharing} provides background on Shamir's secret sharing scheme.

\subsection{Intel SGX}
\label{subsec:SGX}

\noindent\textbf{Enclave Execution.} 
Intel SGX~\cite{sgx} is a set of CPU extensions available on modern Intel processors~\cite{intel_skylake}.
These CPU extensions are designed to provide hardware-protected TEE (or \textit{enclave}) for generic computations. 
Intel SGX associates each enclave process with a CPU-guarded address space (aka enclave memory). 
The CPU prevents any foreign process (i.e., non-enclave process) from accessing the enclave memory. 
The enclave is isolated from other enclaves concurrently running on the same host, from the OS, and from other user processes. 
Intel SGX allows paging for the enclave memory, and the memory pages are encrypted under the processor's key prior to leaving the enclave. Enclaves cannot directly execute OS-provided services such as I/O.
They have to employ OCalls (i.e., calls that are executed by the enclave code to transfer the control to non-enclave code) and ECalls (i.e., API for untrusted applications to transfer control into the enclave) to facilitate those services.

\vspace{2mm}
\noindent\textbf{Attestation.}
Intel SGX's attestation mechanisms~\cite{sgx_remote_attest} enable a validator to verify if an enclave in question is instantiated with the correct code. At the same time, these mechanisms provide means via which the validator and the attesting enclave can establish a secure, authenticated channel to communicate sensitive or private data. 

If the validator is another enclave instantiated on the same platform (or host) with the attesting enclave, it can use {\em local attestation} to ascertain the correct instantiation of the latter. 
More specifically, after  the attesting enclave is initiated, the processor produces its \textit{measurement} (i.e., the hash of its initial state), and then creates a message authentication code (MAC) of such measurement using a key that can only be retrieved by the validating enclave. The
measurement of the attesting enclave and its MAC are sent to the validating enclave for verification.
On the other hand, if the validator is a remote party, the processors signs the measurement with its private key under the Enhance Privacy ID (EPID) scheme~\cite{epid}~\cite{sgx_remote_attest}, generating a {\em remote attestation}. It is worth noting that in the current SGX architecture, the remote party obtaining the attestation cannot complete the verification of such attestation by itself. Instead, it has to rely on the Intel's Attestation Service (IAS) to check if the signature contained in the attestation~\cite{ias} is valid. Once the signature is validated, the validating party checks the measurement value against a known value to make certain of the correct instantiation of the attesting enclave.

\vspace{2mm}
\noindent\textbf{Data sealing.} Enclave memory is volatile. To persist their private state to non-volatile storage, enclaves need to employ data sealing mechanism. Concretely, to seal its private state or data to persistent storage, the enclave first obtains a unique key that is bound to its measurement from the processor. Subsequently, it encrypts the data using the enclave-specific key, and offloads the encrypted data to the storage. This data sealing mechanism guarantees that the sealed data can only be retrieved by the enclave that sealed it. However, data sealing and retrieving is susceptible to rollback attacks wherein an adversary (e.g., the malicious OS) presents the enclave with properly sealed but stale data~\cite{rollback_detection}. We refer readers to~\cite{rote} for defense against such rollback attack.

\subsection{Consensus Protocols}
\label{subsec:consensus}
In a distributed system comprising of independent and mutually distrusting nodes, having the nodes agree on some data that is crucial for the operation of the system is challenging, even more so in the presence of node failures. An extensive body of research, especially on \textit{distributed consensus protocols},  has been dedicated to address a variety of fault tolerance problems in distributed systems~\cite{raft, pbft}. There are two types of failures a node may undergo, namely {crash failure} and {Byzantine failure}.  The former characterizes situation in which a node abruptly stop and does not resume~\cite{raft}, while the latter, which is more disruptive in comparison with the former, observes a faulty node deviating arbitrarily from its expected behaviors. For example, a node experiencing Byzantine failure may equivocate (i.e., sending contradictory messages to other nodes), or it may intentionally delay its activity for any period of time~\cite{pbft}. 

Consensus protocols are designed to achieve \textit{safety} and \textit{liveness} in the presence of failures. Safety necessitates non-faulty nodes to reach an agreement and never return conflicting results for the same query, whereas liveness requires that these nodes eventually agree on a value. The rest of this section focuses on a particular crash failure consensus protocol called Raft~\cite{raft}.

\vspace{2mm}
\noindent{\bf Raft Consensus Protocol.}
Raft assumes a system of $n$ deterministic processes (or nodes), among which at most $f = \frac{n-1}{2}$ could be faulty. A faulty process fails by crashing. Each process stores a log that contains a series of commands, or events of interest. Raft ensures that logs of non-faulty processes contain the same sequence of commands, thereby achieving safety even in asynchronous network (i.e., regardless of timing), and necessarily depends on timing to offer liveness~\cite{FLP_impossibility} (e.g., network is partially synchronous such that messages are delivered within an unknown but finite bound).

Under Raft protocol, nodes assume one of the three roles, namely {\it leader, follower} and {\it candidate}, and time is split into {\it terms} numbered by consecutive integers. 
Each term marks one node as the leader, while other nodes are followers. 
The leader maintain its authority by periodically exchanging heartbeats with all followers.
Should a follower fail to hear from the leader after an \textit{election timeout} period, it shall consider the leader crashed, increase its own term, undertaking the candidate role and triggers an election. The candidate claims itself as a leader once it has collected votes from a majority of nodes. 
More detailed discussion on the leader election and its election criteria can be found in Raft's original paper~\cite{raft}.

Followers respond to requests from the leader and candidate, and stay passive otherwise. 
All commands (e.g., clients' requests) are sent to the leader, and subsequently replicated on the followers.
Given a command, the leader first appends it as a new entry uniquely identified by the leader's \textit{term} and an index to its log. It then announces the entry to all followers. 
Upon receiving a new entry, followers append it to their logs, and acknowledge the receipt with the leader. 
After receiving the acknowledgement from a majority of nodes in the system (i.e., $f+1$ or more nodes), the leader \textit{commits} the entry by executing the command it contains, as well as all preceding entries in its log if they have not been committed. 
The leader keeps track of the highest index committed, and convey this information to the followers in subsequent messages so as to inform the latter on the committed entries.

While Raft assumes crash failure model, one can deploy it in a Byzantine setting through the use of TEE ~\cite{mscoco, AMES}. In particular, adversarial behaviours of the faulty nodes can be restricted by running the codebase of the consensus protocol inside the TEE with attested execution. This effectively transform the Byzantine threat model to crash fault tolerance, in which Raft is applicable.

\subsection{Threshold Secret Sharing}
\label{subsec:secre_sharing}
A threshold secret sharing scheme (e.g., Shamir's Secret Sharing~\cite{shamir_secret_share}) is parameterized by two values, namely $n$ and $t$. The scheme preserves a secret $S$ in a distributed manner. More specifically, the secret $S$ is split into $n$ multiple \textit{shares}. 
To reconstruct the original secret $S$, one would need a minimum number of shares. This number is the threshold, and denoted by $t$. Any set of $t$ or more shares can be used to reconstruct the original secret $S$.
The security of the scheme dictates that no adversary that possesses knowledge of any $t-1$ or fewer shares could determine or reveal any information about $S$. 

Shamir's Secret Sharing~\cite{shamir_secret_share} works as follows. Given a secret $S$, the threshold $t$ and the number of shares $n$, wherein $t, n$ and $S$ are elements in a finite field $\textsl{F}$ of size $\textsl{P}$, $0 < t \leq n < \textsl{P}$; $S < \textsl{P}$ and $\textsl{P}$ is a prime number, one can secret share $S$ as follows. First, let $a_0 = S$, and choose $t-1$ positive integers $a_1, \ldots, a_{t-1}$ at random such that $a_i < P$ $\forall i \in [1, t-1]$. Next, construct a polynomial $f(x) = a_0 + a_1x + a_2x^2 + \ldots + a_{t-1}x^{t-1}$, and pick $n$ points out of form $(j, f(j))$ on $f(x)$'s graph. Every such point (plus the knowledge of the prime number $\textsl{P}$ defining the finite field $\textsl{F}$) constitutes a \textit{share}. Given the knowledge of any $t$ or more such points, one can 
evaluate the coefficients $a_0, a_1, \ldots, a_{t-1}$ of $f(x)$ using interpolation, and obtain the secret $S$ from the constant term $a_0$. 

\section{The Problem}
\label{sec:the_problem}
In this section, we characterize the self-expiring data encapsulation framework we seek, and challenges we must 
overcome in provisioning one. We first discuss a few application scenarios that motivate the needs of self-expiring data capsules in 
Section~\ref{subsec:usecase}. Next, Section~\ref{subsec:prob_overview} gives an overview of the data encapsulation framework we 
study, while  Section~\ref{subsec:system_goal} elaborates on the goals we aim to offer. Subsequently, Section~
\ref{subsec:adv_model} details the adversary model we consider. Finally, we describe challenges faced in designing the self-expiring 
data capsule by surveying candidate approaches and pointing out their limitations in Section~\ref{subsec:challenges}.

\subsection{Use Cases}
\label{subsec:usecase}

There exists physical implementation of expiring/disappearing data object. Xerox PARC~\cite{self_destruct_chip} introduced a chip that self-destructs upon command. 
The chip can be used to store sensitive data such as encryption keys. When there is
a need to dismantle the sensitive data, a command can be issued to the chip, causing it
to physically shatter into smithereens. Once shattered, the chip cannot be reconstructed from the 
smithereens, making the data irretrievable. However, the destruction of the chip requires explicit invocation from the chip owner.

Ephemeral messaging or social media services, such as Snapchat and 
Instagram, feature expiring/disappearing data object in their workflow. 
These applications typically have a default ``expiry condition'' for 
messages sent or stories posted by their users. 
Once such a condition becomes true (e.g., after a 
Snapchat message has been seen, or 24 hours after an Instagram story is posted), 
these platforms render the messages/stories inaccessible to the original intended viewers 
without any action from the users. Nevertheless, it is unclear whether the content (e.g., messages and 
stories) is completely deleted from the cache and storage systems of the service providers.

Other applications could also take advantage 
of the self-expiring data capsule. Consider private and event-specific emails whose 
content is related to a distinct event and should only be available to parties
directly involved with that event. Such emails may cease to have any value to the involved parties after 
the associated event has come to pass. However, email providers (e.g., Gmail or Yahoo) tend to store them 
for an extended period of time (or even indefinitely) on their servers, which gives rise to potential 
privacy risk wherein the content of the email is revealed to unintended parties against the will of 
senders (potentially due to mismanagement of the email providers, or subpoena). For instance, during 
dialogue leading up to the actual incorporation of a company, co-founders or stakeholders may have email 
exchange discussing different proposals with regard to the internal structure of the to be incorporated 
company, and these proposals contain sensitive information that they wish not to disclose to any 
outsider. Once they have reached a consensus and finalized the company structure, the emails that contain 
disregarded proposals no longer have any value, and the stakeholders would prefer all copies of those 
emails to be automatically deleted regardless of where they are cached and stored rather than observing a 
risk of them being disclosed to unintended parties. In such scenario, self-expiring email capsule is 
clearly of great interest.

As yet another example, let us consider a group of credible entities, such as hospitals, which are in possession of 
a large corpus of sensitive and valuable data, in particular protected health information and clinical records of their patients.
These data providers (i.e., hospitals) may be mutually distrustful, and impose different privacy-protecting policies on their data sets.
However, they are willing to allow selected clinical researchers to mine their data sets, for example to facilitate the study of personalized treatment or to evaluate their new findings, so long as the individual records are kept private, and the mining adheres to their privacy-protecting policies. The hospitals may employ privacy-preserving technique (e.g., differential privacy~\cite{differential_privacy}) to sanitize the answers given to queries submitted by the researchers. Without loss of generality, we assume that such privacy-protecting policy/mechanism is associated with a privacy budget, and the hospitals would like to render their data inaccessible to the researchers once the privacy budget has been exhausted. Moreover, the hospitals also want to restrict the type of computations that the researchers can perform on their data set, requiring the researchers to specify the specification of the   functions, and scrutinizing such functions before granting them access to their protected data.

The examples discussed earlier observe a common trait. That is, the data of interest is available for a 
particular set of eligible parties and only for a limited duration until it is expired, subject to a 
predefined access control policy. Once the data is expired, it is no longer accessible by anyone, 
including the previously eligible parties, any entity legitimately or illicitly maintaining or caching 
its copies, and the adversary.

\subsection{Problem Overview}
\label{subsec:prob_overview}
We focus on the deployment setting that comprises the following main parties, namely {\it data owner}, {\it encrypted storage server}, {\it access control committee} and {\it  access requester}. 

\begin{itemize}
 \setlength\itemsep{0.3em}
 
\item \textbf{Data Owners} are in possession of sensitive and valuable data, and willing to allow scrutinized   functions to access their sensitive data. Such accesses must adhere to access control and privacy protection requirements imposed on the sensitive data. Without loss of generality, we assume that the sensitive data is encrypted using semantically secure symmetric-key encryption scheme~\cite{crypto_intro}, and its ciphertext (or capsule) is stored on a public encrypted storage server (discussed below). Access to the data is granted via the delivery of the encryption key. The access control conditions cover the eligibility of the   functions, as well as the expiring condition of the protected data. Once the data is expired, it must be rendered inaccessible to anyone, including the previously eligible   functions. Nonetheless, the data owners do not wish to upkeep the access control management on their own, but rather outsource it to an access control committee (discussed below) that is designed to be  robust against single-point-of-failure issues. 

\item \textbf{Encrypted Storage Server} is a key-value database that stores the encrypted data. The generality of the framework requires that its design is agnostic to the implementation of the encrypted storage server (e.g., RocksDB~\cite{rocksdb}, LevelDB~\cite{leveldb}). Interactions with the  storage  are distilled into two main functions, namely $\texttt{store(k, v)}$ which puts the data $\texttt{v}$ indexed by the key $\texttt{k}$ onto the storage, and $\texttt{retrieve(k)}$ which returns the data indexed by $\texttt{k}$ if it has been stored. 

\item \textbf{Access Control Committee} comprises \textit{independent} and \textit{mutually distrusting} nodes that collectively enforce the access control mechanism dictated by data owners on the data capsules. The access control committee is a unique portal via which eligible   functions can obtain access to the protected data (i.e., its encryption key). Given an access request, the nodes in the committee collectively determine whether the requested data has not been expired, and the requesting function is eligible for the access.  
If the data is available, and the eligibility of the requesting function checks out, the committee securely delivers the encryption key to the requesting function. 
We emphasize that the eligibility in this context encompasses not only the execution logic of the   functions, but also the assurance that they do not leak the sensitive data (in its plaintext form) or its encryption key to any unintended party, and do not persist plaintext copy of the data or the encryption key on its storage.

\item \textbf{Access Requesters} are those that wish to access the protected data using specific functions (or programs). It is required that such  functions are executed inside TEEs, and their measurement is known to the data owners. 
\end{itemize}

\definecolor{airforceblue}{rgb}{0.36, 0.54, 0.66}
\definecolor{aliceblue}{rgb}{0.94, 0.97, 1.0}
\definecolor{apricot}{rgb}{0.98, 0.81, 0.69}

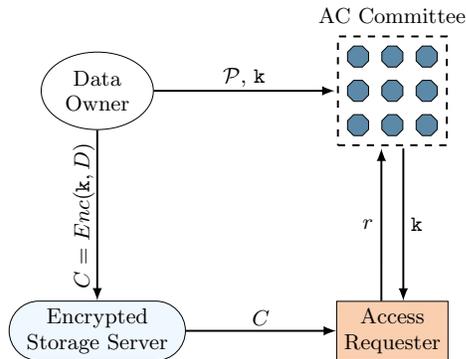
\begin{figure}
\centering    
\resizebox{.36\textwidth}{!}{  
\begin{tikzpicture}[scale=0.4, every text node part/.style={align=center}]  
\node[regular polygon, regular polygon sides = 8, draw = black, fill = airforceblue] (node00){};
\node[regular polygon, regular polygon sides = 8, draw = black, fill = airforceblue, below of = node00, node distance = 0.5cm] (node10){};
\node[regular polygon, regular polygon sides = 8, draw = black, fill = airforceblue, below of = node10, node distance = 0.5cm] (node20){};
\foreach \name / \y in {2/0.5,3/1}
\node [regular polygon, regular polygon sides = 8, draw = black, fill = airforceblue, right of = node00, node distance= \y cm] (node0\name) {};
\foreach \name / \y in {2/0.5,3/1}
\node [regular polygon, regular polygon sides = 8, draw = black, fill = airforceblue, right of = node10, node distance= \y cm] (node1\name) {};
\foreach \name / \y in {2/0.5,3/1}
\node [regular polygon, regular polygon sides = 8, draw = black, fill = airforceblue, right of = node20, node distance= \y cm] (node2\name) {};

\draw[thick,dashed] ($(node00.north west)+(-0.45,0.45)$)  rectangle ($(node23.south east)+(0.45,-0.45)$);

\node[ellipse, draw = black, left of = node10, node distance = 3.8cm] (data_owner) {Data \\ Owner};

\node [rounded rectangle, draw = black, below of = data_owner, node distance = 3.5cm, fill = aliceblue] (ESS) {Encrypted \\ Storage Server};

\node [rectangle, draw = black, right of = ESS, node distance = 4.3cm, fill=apricot] (requester) {Access \\ Requester};

\node [draw = none, above of = node02, node distance = 0.6cm](){AC Committee};

\draw[-latex, thick] (data_owner.east) -- node[yshift = 0.6em, inner sep =1.5pt] {$\mathcal{P}$, $\mathtt{k}$}  ($(node10.west)+(-0.45,0)$);
\draw[-latex, thick] (data_owner.south) -- node[rotate = 90, yshift = 0.6em]{$C = Enc(\mathtt{k}, D)$} (ESS.north);
\draw[latex-, thick] (requester.west) -- node[yshift = 0.6em]{$C$} (ESS.east);
\draw[-latex, thick] ($(requester.north)+(-0.4,0)$) -- node[xshift=-0.6em]{$r$}  ($(node22.south)+(-0.4,-0.45)$);
\draw[latex-, thick] ($(requester.north)+(0.4,0)$) -- node[xshift=0.6em]{$\mathtt{k}$}  ($(node22.south)+(0.4,-0.45)$);

\end{tikzpicture}
}
\caption{An overview of the interplay between the parties. The data owner encrypts her data $D$ under key $\mathtt	{k}$, obtaining the ciphertext $C$. She then entrusts the key $\mathtt{k}$ and the access control policy $\mathcal{P}$ to the access control (AC) committee, and $C$ to the encrypted storage server. The access requester submits a request $r$ to the AC committee, and obtains the $\mathtt{k}$ if his function is eligible to access $D$ per the policy $\mathcal{P}$.}
\label{fig:Overview}
\end{figure}

\subsection{System Goals}
\label{subsec:system_goal}
This section enumerates the goals and properties that a self-expiring data encapsulation framework must offer. The resulting construction is expected to enhance privacy protection of numerous applications that involve digital content without requiring explicit intervention of the data owners in the upkeep of the access management.

\begin{itemize}
 \setlength\itemsep{0.3em}
\item \textbf{Generic expiring condition}: The framework must support generic expiring conditions (e.g., beyond simple timeout), giving the data owners flexibility in defining the expiration criteria of their protected data.

\item \textbf{Unavailability after expiration}:  An expired data capsule can no longer be decrypted, and the protected data remains inaccessible to anyone, including the previously eligible access requester and their scrutinized   functions as well as entities legitimately or illicitly maintaining or caching copies of the data capsule regardless of when they obtain such copies.

\item \textbf{Rigid access eligibility}: Prior to the data capsule's expiration, the protected data can be made available to eligible access requesters via their properly scrutinized   functions. It is important to stress that the access eligibility is determined based on not only  the function's execution logic, but also the assertion that such execution does not reveal the plaintext copy of the protected data or its encryption key to any unintended party, nor persist them in its storage.

\item \textbf{Autonomy}: The access control operations (e.g., eligibility checking, data capsule expiration) should be performed in an autonomous yet consistent manner by the nodes in the committee. No explicit action from the owners or parties who cache and archive the data capsule are required.

\item \textbf{No-added attack vector}: The self-expiring data capsule must not expose any new attack vector on the protected data in comparison to the baseline solution wherein the data owners keep their data in private, manually authorize every access request, and communicate the requested data to the eligible requester.

\end{itemize}

Let us revisit the example of protected health information and clinical records mentioned in Section~\ref{subsec:usecase}.
Under the setting described in Section~\ref{subsec:prob_overview}, the hospitals play the role of data owners, 
whereas the researchers are access requesters. Each hospital may encrypt its data corpus with a unique 
encryption key, and may not share this key with other hospitals. The encrypted health information and clinical 
records can be placed on a publicly accessible storage. The security of the encryption scheme in use assures that 
no party can retrieve the clinical data in plaintext without first obtaining the corresponding encryption key. Access 
to the clinical data is granted via the delivery of the encryption key. To ensure that the researchers will not deviate 
from the intended and authorized computations to be conducted on the clinical data, nor illicitly persist a copy of 
the clinical data on their storage, the handling of the encryption keys and processing of the protected data must 
be carried out exclusively inside TEEs with attested execution.

In the absence of a secure self-expiring data encapsulation framework, both the data owners and the access requesters have to engage in involved and laborious processes. The hospitals have to check the eligibility of each researcher requesting for data access and the availability status of the requested data with respect to its privacy budget. Furthermore, they also need to verify the specification of the   function as well as the correct instantiation of the corresponding TEE. Besides, since different hospital encrypts their data using different encryption key and observes different privacy budget, a researcher who wishes to source data from multiple hospitals has to send separate requests to the relevant hospitals, and in some situation, has to await the coordination of the mutually distrusting hospital in granting access to the requested data. This onerous practice clearly poses too much burden on both parties, and hinders the scalability of the system.

\subsection{Assumptions and Adversary Model}
\label{subsec:adv_model}
In this section, we specify the trust assumptions as well as the adversary model we consider in designing a self-expiring data encapsulation framework that attains the desired properties and goals described in the previous section. 

\vspace{2mm}
\noindent\textbf{TEE Assumptions.} As mentioned earlier, we require the access requesters'  functions to be run inside TEEs with attested execution. 
Our design assumes that the mechanism which is used to provision the TEE (e.g., trusted processors such as Intel SGX~\cite{sgx} or Sanctum~\cite{sanctum} or Keystone~\cite{keystone}) is implemented and manufactured properly. 
Although our work adopts Intel SGX processors to provision the TEEs, we remark that our design is also compatible with other TEE instantiations (e.g., TrustZone~\cite{trustzone}, Sanctum~\cite{sanctum}). 
While SGX enclaves admit side-channel and rollback attacks~\cite{controlled_channel, rollback_detection}, we note that both software and hardware techniques to mitigate these attacks are presented in the literature~\cite{STC, rote}. Incorporating defences against these attacks is orthogonal to our work.

\vspace{2mm}
\noindent\textbf{Access Control Committee Assumptions.} Nodes in the access control committee are independent and mutually distrusting. Although the nodes may not trust each other, they trust the codebase implementing the committees' functionality, which can be formally vetted and verified. 
In order to prevent a malicious node to deviate from the trusted codebase, we require every node in the committee to run such a codebase inside a TEE featuring attested execution, for instance enclaves provisioned by commodity Intel SGX-enabled processors~\cite{intel_skylake}. 
The nodes' pairwise communication is conducted via an authenticated and reliable point-to-point communication channel, and no message is dropped. In the absence of the adversary (discussed below), these channels are \textit{synchronous} (i.e., all messages are delivered within a finite delay $\Delta$ known a-priori). 
Let us denote by $n$ the number of nodes in the committee. A node is considered \textit{faulty} if it crashes, or it cannot communicate with a quorum of $\lfloor \frac{n}{2} \rfloor$ nodes in a synchronous fashion. 
Let $f$ be the number of faulty nodes in the committee, we require $f \leq \lfloor \frac{n-1}{2} \rfloor$ in order for the committee's operations to remain intact and secure.

\vspace{2mm}
\noindent\textbf{Threat Model.}
The main adversarial goal is to bypass the protection mechanisms of the framework, gaining illegitimate access to the plaintext copy of the protected data even though it does not meet the access control conditions. We study an  adversary that can corrupt upto $\lfloor \frac{n-1}{2} \rfloor$ nodes in the  committee. It has full control over the compromised nodes. It can corrupt or schedule all non-enclave processes, including the node's operating system (OS), and access their memory content. Moreover, it can tamper with data persisted on the compromised node's storage. It can intercept, read, modify, reorder and delay messages sent from and to the compromised nodes. The adversary can also pose as an access requester in an attempt to leak the plaintext copy of the protected data. Alternatively, it can compromise an otherwise benign access requester in the same way it can compromise a node in the access control committee. Finally, since the encrypted storage server is publicly accessible, we assume that the adversary can retrieve any encrypted data stored there.

Nonetheless, the adversary cannot violate protections mechanisms provided by the SGX processors. More specifically, it is unable to access runtime memory of an enclave, or leak confidential information protected by the enclave execution. The adversary does not have any access to the SGX processor's private keys which are utilized in attestation and data sealing~\cite{sgx}. The adversary is computationally bounded, and it cannot break standard cryptographic assumptions.

We assume the application codes running inside the TEEs (e.g., the codebase implementing functional logic of the access control committee, access requesters'  functions) can be formally vetted and therefore trusted. We leave side-channel attacks against the enclave execution~\cite{controlled_channel} and Denial-of-Service attacks against the system out of scope.

\subsection{Challenges}
\label{subsec:challenges}
Designing a data encapsulation framework that achieves all the goals and features discussed in Section~\ref{subsec:system_goal} is 
technically challenging. Let us consider a  naive approach in which the data owners manage access to their  sensitive data by themselves. While it can support generic expiring condition, ensure unavailability of the protected data after 
expiration, and guarantee rigid access eligibility, it requires intricate and onerous involvement of the data owners. One may attempt to 
unburden the data owners by introducing a dedicated server that handles the key and access management on the data owners behalf. 
Nonetheless, this solution suffers from single-point-of-failure.

The dedicated server features enclave execution, and maintains the encryption keys of data owners inside the enclave memory or on the server's persistent storage using data sealing~\cite{sgx}. Upon receiving 
an access request, it will first check the availability status of the requested data, and the eligibility of the requesting function (e.g., 
via remote attestation). If all the requirements check out, the server securely delivers the relevant encryption keys to the the TEE of the function, thereby granting it access to the protected data. The security of this approach essentially depends 
upon the capability of the dedicated server (i) to prxeserve the confidentiality of the encryption keys, (ii) to ensure the freshness of the 
availability status of the protected data, and (iii) securing the execution integrity of the access control management logic. While 
enclave execution provides strong protection to (iii), attacks that compromise confidentiality of enclave execution~
\cite{controlled_channel} threaten to breach (i), and rollback attacks~\cite{rollback_detection} imperils (ii). In case the 
dedicated server crashes indefinitely, the encryption keys are lost, causing the data capsules to expire prematurely. 

One may battle the single-point-of-failure using replicated state machine. Instead of 
having one dedicated server, we can rely on a committee of independent nodes to 
collectively implement such logic. For instance, the availability status of the protected data is replicated across all the nodes in the 
committee, and determining whether the requested data has been expired requires a consensus among the nodes. This approach 
strengthens (ii), for it is arguably reasonable to assume that launching a coordinated rollback attack on a set of independent node is 
much harder than perpetrating the attack on a single node. Nonetheless, care must be taken to assure (i), since merely replicating the 
encryption keys does not mitigate the threats posed by attacks against confidentiality of enclave 
execution. As soon as the adversary successfully extracts encryption keys from one of the nodes, the security of the framework 
collapses.

\section{The \codename\ Design}
\label{sec:design}
We now present our self-expiring data encapsulation framework, which we name \codename. We first provide key insights behind the design of \codename\ in Section~\ref{subsec:insight}. Next, we examine implementation alternatives, and analyse their characteristics. 
Finally, we detail the workflow of \codename.
We note that while our system adopts Intel SGX to provision TEEs (or enclaves) with attested execution, the underlying insight of \codename\ is also applicable to other TEE instantiations (e.g., Keystone~\cite{keystone}, Sanctum~\cite{sanctum}).

\subsection{Key Insights}
\label{subsec:insight}
The security of our framework dictates that access to the protected data is only granted to scrutinized   functions of eligible access requesters, that it is not leaked to any unintended party, and that it becomes inaccessible to anyone, including the previously eligible access requesters after its expiration. In order to guarantee these properties, \codename\ needs to uphold the following two invariants:

\begin{enumerate}
 \setlength\itemsep{0.3em}
\item \textit{Confidentiality of encryption keys}: The encryption keys that can be used to decrypt the protected data (i.e., decapsulate the data capsule) must be kept private (except for the data owner himself) at all time. It is only revealed to the  eligible function once the access is granted. Further, it is important to ensure that such  function, once granted the encryption key, can only use it inside the TEE, and does not persist the key on its storage. Once the data capsule expires, the corresponding encryption key must be eradicated.  

\item \textit{Legitimacy of data accesses}: Without loss of generality, we can assume that each data capsule is associated with an access log. An access requester and her  function, if admitted to decapsulate the data capsule and consume the protected data, must be first recorded in its access log. In another word, the approval of an access request is indicated by its corresponding entry in the log. The expiring condition, and thereby the availability status of a data capsule can be defined and determined based on the access log. Consequently, in order to ensure legitimacy of data accesses and enforce proper expiration of the protected data, it is critical to ensure the integrity and tamper resistance of the access logs in the presence of adversary. 

\end{enumerate}

While it is convenient to have a dedicated server manages the two invariants mentioned above, this approach is clearly susceptible to single point of failures. A proven strategy to sidestep  such failures is to embrace decentralization. It is worthy to note that decentralization goes beyond replication. As an example, splitting a secret into shares such that a subset of these shares can be used to reconstruct a secret, and depositing them to independent nodes offers better protection to the confidentiality of the secret in comparison with storing the said secret at one node. On the other hand, replicating such secret at different independent nodes does not offer any stronger confidentiality protection compared to having only one node storing the secret, for compromising a single node among them is sufficient to extract the secret. 

As indicated above, a natural approach to decentralise a secret is to adopt a secret-sharing scheme~\cite{shamir_secret_share, churp}. Under a $(t,n)$-secret sharing scheme, a secret $S$ is split into $n$ shares given to $n$ independent nodes. Any subset of $t$ or more shares can be used to reconstruct the secret, while individual shares do not reveal any information about $S$. The adversary must corrupt at least $t$ nodes and obtains their shares in order to leak the secret. The secret-sharing also enhances robustness of the system against  secret loss, for $S$ is only rendered unattainable  when more than $n-t$ shares are lost. Our design takes advantage of secret sharing to strengthen the protection of the encryption keys.

Besides preserving confidentiality of encryption keys, a self-expiring data encapsulation framework needs also to ensure legitimacy of data accesses. In designing \codename, we observe an interesting analogy between a data capsule in our context and a coin (e.g., Bitcoin) in the crypto-currency context~\cite{btc_origin}. More specifically, while we want to guarantee that only eligible  function can access the protected data, and that access is only possible while the data capsule has not yet expired, a crypto-currency platform needs to ensure that a coin can only be spent by its owner (e.g., to whom the coin has been sent to previously), and the owner can only spend that coin once. Under our framework, all accesses to the protected data are recorded in its associated access log, and the expiration status of the said data can be determined based on the content of the access log. Likewise, all transactions that spend coins in the crypto-currency platform are recorded in a ledger, and one can rely on such a ledger to prevent double-spending.  \codename\ shares with the crypto-currency platform a necessity of securing the integrity and tamper resistance of an append-only sequence of data records. This very analogy suggests replicated state machine~\cite{RSM} as a natural solution  to safeguard the validity and incorruptibility of the access logs in our framework. That is, the access logs are replicated across independent nodes which run a distributed consensus protocol to assure that  their local copies of the logs converge even in the presence of the adversary or individual node failures.

\subsection{Implementation Consideration}
\label{subsec:design_rationale}
We have pointed out the two key invariants that \codename\ must afford, and described how
decentralization can be adopted to implement them. 
Nevertheless, there remains a critical implementation choice to be determined. 
The first alternative is to  have a single committee of independent nodes handling both properties. 
On the other hand, one could employ two separate, non-overlapping committees, 
each of which safeguards one invariant, provided that their operations are properly coordinated.
In what follows, we sketch an outline of each design choice and explore its implications.

\vspace{2mm}
\noindent\textbf{Two separate, non-overlapping committees.} This design choice requires the involvement of two separate sets of independent nodes. The first, which we call encryption-key-committee (or $\mathtt{EKComm}$), is tasked to protect the confidentiality of the encryption keys via the use of secret-sharing. The second set of nodes, dubbed access-log-committee (or $\mathtt{ALComm}$) is tasked to safeguard the validity and incorruptibility of the access logs associated with the data capsules. It is needless to say that nodes in $\mathtt{EKComm}$ does not have permission to write to the access logs, and nodes in $\mathtt{ALComm}$ does not have any knowledge of the secret keys or their shares.

Having two separate committees allows greater flexibility in configuring system and security parameters. For instance, $\mathtt{EKComm}$ may comprise a large number of commodity nodes, and the threshold used in the secret sharing is high (e.g., $t = \frac{2}{3} n$). Larger $\mathtt{EKComm}$ and higher threshold means the adversary has to corrupt more nodes in order to extract the encryption keys. On the other hand, $\mathtt{ALComm}$ can be configured to consist of only a small number of nodes with defense-in-depth mechanisms~\cite{fraser2000hardening, stytz2004considering} in place to better safeguard them against adversarial compromise. A small $\mathtt{ALComm}$ is expected to yield higher throughput (i.e., the number of access request the committee can process per unit of time), for consensus protocols designed for permissioned settings often favor small committee size~\cite{sigmod_sharding}. One may also argue that this design choice adheres to the principle of privilege separation - a technique in which different parts of the system are limited to specific privileges and tasks.

Nonetheless, the separation of $\mathtt{EKComm}$ and $\mathtt{ALComm}$ presents challenges in coordinating the interactions between the access requester $\mathtt{R}$ and the two committees. Recall that in \codename\, the approval of an access request is indicated by its corresponding entry in the access log, and the access is granted via the delivery of the encryption key. That is, granting access to a data capsule is essentially a distributed and atomic transaction~\cite{nonblocking_commit}. In our context, this comprises two  indivisible operations, one concerning $\mathtt{EKComm}$ while the other involving $\mathtt{ALComm}$. These two operations are distributed in a sense that two committees responsible to handle them are separate and independent.

A standard approach to ensure atomicity of a distributed transaction is to adopt an atomic commitment protocol, such as two-phase commit protocol (2PC)~\cite{concurrency_control_book}. In a classic 2PC protocol, there is a coordinator and a set of participants. The protocol starts by the coordinator querying all participants if they can commit the transaction. Next, the coordinator broadcasts a $\texttt{commit}$ message if it receives a vote $\texttt{yes}$ from every participant, or an $\texttt{abort}$ message otherwise. When the participants receive the $\texttt{commit}$ message, they complete the transaction locally and responds the coordinator with an acknowledgement. Alternatively, if they receive the $\texttt{abort}$ message, they will undo all operations related to the transaction in question, and revert to the initial state (i.e., before they receive query to commit from the coordinator).

One may apply 2PC protocol on the two committees $\mathtt{EKComm}$ and $\mathtt{ALComm}$ as follows. Let the former  serve as a coordinator, while the latter be the participant. 
The access requester $\mathtt{R}$ sends the request to $\mathtt{EKComm}$, which then queries $\mathtt{ALComm}$ for the eligibility of the request. 
If $\mathtt{ALComm}$ replies $\mathtt{EKComm}$ with a vote $\texttt{yes}$, $\mathtt{EKComm}$ reconstructs the secret key, and sends a $\texttt{commit}$ message to $\mathtt{ALComm}$. 
Upon receiving the $\texttt{commit}$ message, $\mathtt{ALComm}$ writes the entry corresponding to the request to the access log, and replies $\mathtt{EKComm}$ with an acknowledgement. 
Upon receiving the acknowledgement, $\mathtt{EKComm}$ delivers the secret key to TEE of the requesting function.  

The description above, however, oversimplifies many subtle details. 
First, since $\mathtt{EKComm}$ and $\mathtt{ALComm}$  are collections of nodes, operations such as $\mathtt{ALComm}$ replying to the query to commit from $\mathtt{EKComm}$, or $\mathtt{EKComm}$ sending a $\texttt{commit}$ message to $\mathtt{ALComm}$ entail nodes in the committees to complete one run of the consensus protocol. 
Secondly, committee-to-committee communication poses heavy stress on both the network and the nodes when the committee sizes are large. 
As an alternative, one may also designate the access requester $\mathtt{R}$ as the coordinator, and have both committees serve as participants. 
Nonetheless, this is not ideal either, for the access requester may fail permanently during the protocol execution, leaving one or both of the committees blocked indefinitely~\cite{sigmod_sharding}.

Another possible workflow is to let the access requester $\mathtt{R}$ interact with $\mathtt{EKComm}$ and $\mathtt{ALComm}$ in two stages. In the first stage, a request is submitted to $\mathtt{ALComm}$, which then verifies the eligibility of the request. 
If the access can be granted, $\mathtt{ALComm}$ confers the requester $\mathtt{R}$'s  function an access token  $\texttt{T}$. 
$\mathtt{R}$ then presents $\texttt{T}$ to $\mathtt{EKComm}$ in order for its  function to obtain the relevant secret key. 
The reconstruction of the secret key from the shares can be done by $\mathtt{EKComm}$, or it shares can be delivered to the  function of $\mathtt{R}$ and reconstructed therein. 
Regardless of where the key reconstruction is carried out, it is important to prevent $\mathtt{R}$ from double-spend the access token. 
This would requires nodes in $\mathtt{EKComm}$  to keep a ledger of used access-tokens, and their ledgers must converge in the presence of the adversary or individual node failures. 
In other words,  $\mathtt{EKComm}$ needs to adopt consensus protocol to keep the nodes' local copies of the ledger consistent. It is also worth mentioning that if $\mathtt{R}$ crashes right after obtaining the access token  $\texttt{T}$ (i.e., the access request is approved but $\mathtt{R}$ never actually decapsulates the data capsule), the access log maintained by $\mathtt{ALComm}$ may not reflect the actual usage of the corresponding data capsule, which may leads to unjust expiration.

\vspace{2mm}
\noindent\textbf{One committee.} We now explore a design choice in which a single committee, called unified access-control-committee (or $\mathtt{UAComm}$), is tasked to uphold both invariants of \codename\  (i.e., confidentiality of encryption keys and legitimacy of data accesses). 
As an overview, given a data capsule $\mathtt{DC}$ with encryption key $\mathtt{k}$, each node in $\mathtt{UAComm}$ keeps a share of $\mathtt{k}$ and a replicated access log associated with $\mathtt{DC}$. 
When an access requester $\mathtt{R}$ submits a request to access $\mathtt{DC}$, nodes in $\mathtt{UAComm}$  collectively determine the eligibility of $\mathtt{R}$'s  function, and the current expiration status of $\mathtt{DC}$. 
If they reach a consensus that the access can be granted, they pool their shares together, reconstruct $\mathtt{k}$, and securely deliver it to the TEE of the requesting  function. 

Compared to the two-committee design choice, utilizing a unified access control committee arguably puts more constraints on system and security parameter configurations. 
In particular, the threshold $t$ of the secret-sharing scheme is set in accordance with the fault-tolerance threshold of the consensus protocol in use. 
In addition, the choice of the committee size becomes more delicate. A small committee size enables $\mathtt{UAComm}$ to afford high throughput, but lowers the robustness of the system against attack that aims to illicitly extract the encryption keys. 
In contrast, a large $\mathtt{UAComm}$ makes it more difficult for the adversary to obtain the secret keys (as it has to compromise more nodes in order to obtained the required shares), but may lessen the the committee's throughput due to higher communication overhead~\cite{sigmod_sharding}.

Handling both the approval of the access request (i.e., its corresponding entry is written into the access log of the data capsule in question), and the reconstruction as well as the delivery of the relevant encryption key, $\mathtt{UAComm}$ alleviates the need of coordinating the interplay between these two operations, naturally ensuring atomicity of access granting.
That is, the access log associated with a data capsule $\mathtt{DC}$ correctly reflects the sequence of accesses that have been granted (i.e.,  functions to which the encryption key has been delivered)\footnote{There exists a possibility wherein the  function crashes immediately after receiving the encryption key, and never actually decapsulates $\mathtt{DC}$.
 Nonetheless, this is beyond the scope of \codename, which concerns only the granting of the accesses, and does not attempt to check the actual consumption of the relevant protected data by the approved  function.}, ensuring just expiration of $\mathtt{DC}$. 
 Moreover, this implementation requires $\mathtt{UAComm}$ to undertake only one  consensus protocol run per each access request  (we shall elaborate in the next Section), which is more efficient and incurs lower latency than the 2PC-based coordination between $\mathtt{EKComm}$ and $\mathtt{ALComm}$ described in the design choice involving two separate, non-overlapping committees.

\vspace{2mm}
\noindent\textbf{Overlapping committees.} Avid readers should have thought of yet another implementation approach wherein the access control committee comprises a large number of nodes, but not all of them are responsible for handling the data capsules' access logs. 
In particular, given a data capsule $\mathtt{DC}$ with encryption key $\mathtt{k}$, each node in the committee keeps a single share of $\mathtt{k}$, but only a select few nodes are tasked to maintain  replicated copies of $\mathtt{DC}$'s access log. 
In a sense, this is equivalent to an implementation that engages two committees, each of which defends one invariant, 
but one committee is a strict subset of  the other. We abuse the notation and denote these two committees by $\mathtt{EKComm}'$ and $\mathtt{ALComm}'$, then  $\mathtt{ALComm}' \subset \mathtt{EKComm}'$. More generally, let us  consider a settings in which $\mathtt{ALComm}'$ may not be a subset of $\mathtt{EKComm}'$, but there exists a non-empty subset of nodes that belong to both $\mathtt{ALComm}'$ and $\mathtt{EKComm}'$ ($\mathtt{ALComm}' \cap \mathtt{EKComm}' \neq \emptyset$).

Similar to the approach utilizing two non-overlapping committees, this setting allows flexibility in configuring system and security parameters. 
That is, one can configure different committee size for $\mathtt{EKComm}'$ and $\mathtt{ALComm}'$, with large $\mathtt{EKComm}'$ enhancing the robustness of the committee in safeguarding the confidentiality of encryption keys, and small $\mathtt{ALComm}'$ yielding high throughput in processing processing  the replicated access logs. 
More interestingly, the intersection of $\mathtt{EKComm}'$ and $\mathtt{ALComm}'$ eases the coordination of their operations with respect to an access request. 
In fact, we believe that it is possible to extend the protocol design constructed for $\mathtt{UAComm}$ (i.e.,  the approval of the access request  and the reconstruction as well as the delivery of the its encryption key to the approved  function are merged into a single protocol run) to support the interactions between $\mathtt{EKComm}'$ and $\mathtt{ALComm}'$, albeit more protocol parameters and convoluted protocol details. 
However, the aforementioned flexibility in system configurations and ease of coordination come at a price of complications in defining the threat model governing $\mathtt{EKComm}'$ and $\mathtt{ALComm}'$, and reasoning  about the security of the system with respect to such a threat model. This is so because faulty nodes belonging to  the intersection of $\mathtt{EKComm}'$ and $\mathtt{ALComm}'$ may entail different security implications in comparison with those that do not belong to the intersection. We leave this exploration to future work.

\subsection{Protocol Details}
\label{subsec:workflow}
Based on the above discussion of advantages and disadvantages of different implementation approaches, we decide to leverage a single access control committee (denoted by $\mathtt{UAComm}$) in the implementation of \codename. The committee  $\mathtt{UAComm}$ consists of $n$ independent nodes $\langle N_1, N_2, \ldots, N_n$. Each node in $\mathtt{UAComm}$ is equipped with an Intel SGX-enabled processor, and their execution is protected by the SGX enclave~\cite{sgx}. 
Our threat model (discussed in Section~\ref{subsec:adv_model} assumes that upto $ \lfloor \frac{n-1}{2} \rfloor$ nodes in $\mathtt{UAComm}$ can be faulty or compromised. A node fails by crashing. The adversary has full control over the compromised nodes, but it cannot break SGX protection mechanisms and standard cryptographic assumptions.
In the following, we present the workflow of \codename, focusing on three key protocols, namely $\textproc{Encapsulate}$ by the data owner, $\textproc{RequestAccess}$ by the access requester, and $\textproc{ProcessRequest}$ by $\mathtt{UAComm}$.

\vspace{2mm}
\noindent\textbf{$\textproc{Encapsulate}$.}
To encapsulate a data object $D$, the data owner $D_{owner}$ follows these steps below:

\begin{enumerate}
 \setlength\itemsep{0.5em}
\item $D_{owner}$ picks an encryption key $\mathtt{k}$ uniformly at random.

\item $D_{owner}$ encrypts $D$ with $\mathtt{k}$ using a semantically secure symmetric-key encryption scheme~\cite{crypto_intro}, obtaining a ciphertext $C$.

\item $D_{owner}$ uses  $(n,t)$ threshold secret sharing scheme~\cite{shamir_secret_share} to split the encryption key $\mathtt{k}$ into $n$ shares ($\mathtt{k}_1$, $\mathtt{k}_2$,$\ldots$, $\mathtt{k}_n$), such that any set of $t=\lfloor \frac{n+1}{2} \rfloor$ shares can be used to reconstruct $\mathtt{k}$. 

\item $D_{owner}$ defines an access policy $\mathcal{P}$ associated with the ciphertext $C$. 
$\mathcal{P}$ specifies the eligibility of the access request and the expiration condition of $C$. 
The data capsule $\mathtt{DC}$ of the data $D$ comprises $\langle C, \mathcal{P}, t \rangle$, and is uniquely identifiable by the file handle $\mathtt{DC}_{ID}$. 

\item $D_{owner}$ entrusts $\mathtt{DC}$ to the encrypted storage server, indexing it with $\mathtt{DC}_{ID}$ (i.e., $\texttt{store}(\mathtt{DC}_{ID}, \mathtt{DC})$).

\item $D_{owner}$  sends $\langle \mathtt{DC}_{ID}, \mathtt{k}_i, \mathcal{P} \rangle$ to node $N_i$ in the access committee $\mathtt{UAComm}$ ($1 \leq i \leq n$) via a secure and authenticated channel.

\end{enumerate}

\vspace{2mm}
\noindent\textbf{$\textproc{RequestAccess}$.}
When an access requester $\mathtt{R}$ wishes to access the data capsule $\mathtt{DC}$ using a  function $\cal{F}$, she needs to perform the following steps:

\begin{enumerate}
\setlength\itemsep{0.5em}
\item $\mathtt{R}$ instantiates the enclave that hosts $\cal{F}$'s execution. 
The enclave instantiation generates a public-private key pair ($\mathrm{pk}_{\cal{F}}, \mathrm{sk}_{\cal{F}}$) uniformly at random, which can be used by a remote party to establish a secure and authenticated communication channel with $\cal{F}$ enclave. 

\item $\mathtt{R}$ obtains  $\cal{F}$'s remote attestation $\pi_{\cal{F}}$ = $\langle M_{\cal{F}}, \mathrm{pk}_{\cal{F}} \rangle_{\sigma_{TEE}}$ from the trusted processor.  ${M}_{\cal{F}}$ is the enclave's measurement, and $\sigma_{TEE}$ is a group signature signed by the processor's private key under the EPID scheme~\cite{epid}\footnote{At the time of this writing, Intel attestation mechanism is designed such that the only party that can verify $\pi_{\cal{F}}$ is the Intel Attestation Service (IAS) acting as group manager. }. 

\item $\mathtt{R}$ requests the Intel Attestation Service to verify $\pi_{\cal{F}}$, retrieving  $\texttt{Cert}_{\mathcal{F}} = \langle \pi_{\cal{F}}, \texttt{valid} \rangle_{\sigma_{IAS}}$ as a response. $\texttt{Cert}_{\mathcal{F}}$ is a publicly verifiable certificate that proves the correct instantiation of $\cal{F}$ enclave. 

\item $\mathtt{R}$ sends a request of form $r = \langle \mathtt{DC}_{ID}, \texttt{Cert}_{\mathcal{F}} \rangle$ to $\mathtt{UAComm}$ (e.g., to the current leader of $\mathtt{UAComm}$), and awaits for the approval and the delivery of $\mathtt{DC}$'s encryption key into the enclave $\mathcal{F}$.

\end{enumerate}

\vspace{2mm}
\noindent\textbf{$\textproc{ProcessRequest}$.}
For a data capsule $\mathtt{DC}$ uniquely identifiable by $\mathtt{DC}_{ID}$, each node in $\mathtt{UAComm}$ maintains a replicated access log that records a sequence of access requester and her  function admitted to decapsulate $\mathtt{DC}$. The availability status of $\mathtt{DC}$ can be determined based on $\mathcal{P}$ and the access log. Therefore, a critical factor of $\textproc{ProcessRequest}$ is to ensure replicated access logs maintained by the independent nodes converge. 
\codename\ leverages a consensus protocol named Raft~\cite{raft} to ensure this convergence.

Raft is designed for crash failure model, whereas our threat model assumes the adversary has full control over compromised nodes. In order to extend Raft to our setting, we require that each node in $\mathtt{UAComm}$ runs $\textproc{ProcessRequest}$ inside the enclave with attested execution. 
This effectively restricts adversarial behaviours of the faulty nodes, thereby reducing the threat model from Byzantine fault tolerance to crash fault tolerance, to which Raft applies. 

Following Raft, one node in $\mathtt{UAComm}$ is elected as a \textit{leader}\footnote{When the leader fails, it is replaced by a leader election protocol. We refer readers to~\cite{raft} for details on leader election and its election criteria.}, while the others are \textit{followers}. 
For clarity, let us denote the leader as $L$. 
Upon receiving a request $r = \langle \mathtt{DC}_{ID}, \texttt{Cert}_{\mathcal{F}} \rangle$ from an access requester, $L$ coordinates the processing of the request as follows:

\begin{enumerate}
\setlength\itemsep{0.5em}

\item  $L$ first verifies if $\texttt{Cert}_{\mathcal{F}}$ is indeed authenticated by the IAS, if $\mathcal{F}$ is eligible to access the data capsule $\mathtt{DC}$ based on $\mathtt{DC}$'s access policy $\mathcal{P}$, and if $\mathtt{DC}$ is still available. If all these conditions check out, $L$ proceeds to the next step. Otherwise, it responds the access requester with $\bot$, indicating access denial. 

\item $L$ puts an entry $\langle \mathtt{DC}_{ID}, \texttt{Cert}_{\mathcal{F}} \rangle$ to its cache, and broadcasts $\langle \mathtt{DC}_{ID}, \texttt{Cert}_{\mathcal{F}} \rangle$ to all the followers.

\item Upon receiving $\langle \mathtt{DC}_{ID}, \texttt{Cert}_{\mathcal{F}} \rangle$, a follower $N_i$ performs the same checks that $L$ has done in step (1). If all the conditions check out, $N_i$ responds to $L$ with an acknowledgement and the key share it keeps (i.e., $\langle ack_i, \mathtt{DC}_{ID}, \mathtt{k}_i \rangle$). Next, it puts $\langle \mathtt{DC}_{ID}, \texttt{Cert}_{\mathcal{F}} \rangle$ to its cache.

\item Once the leader $L$ has confirmed that $\langle \mathtt{DC}_{ID}, \texttt{Cert}_{\mathcal{F}} \rangle$ has been replicated on a majority of the nodes in $\mathtt{UAComm}$, its adds $\texttt{Cert}_{\mathcal{F}}$ to the access log associated with the data capsule $\mathtt{DC}$.
Once it has obtained sufficient number of shares (i.e., $t$) to reconstruct the encryption key $\mathtt{k}$, it reproduces the encryption key $\mathtt{k}$.

\item $L$ then establishes a secure and authenticated communication channel to $\mathcal{F}$ enclave (using $\mathrm{pk}_{\cal{F}}$ found in $\texttt{Cert}_{\mathcal{F}}$), and delivers the encryption key $\mathtt{k}$ to the latter. By this time, $L$ has committed the access granting. 

\item If granting $\cal{F}$ access to leads to the expiration of $\mathtt{DC}$, $L$ discards its the encryption key share it keeps. The leader announces the commit of the access granting in the next message it exchanges with the followers. 

\item Once a follower $N_i$ sees the commit, it adds $\texttt{Cert}_{\mathcal{F}}$ to its access log associated with the data capsule $\mathtt{DC}$. If this leads to the expiration of $\mathtt{DC}$, $N_i$ discards the share $\mathtt{k}_i$ assigned to it. 

\end{enumerate}

\section{Security Arguments}
Since we intend our contributions to be pragmatic as opposed to theoretical, we provide in this section informal security arguments of 
\codename. These arguments justify how \codename\ attains the system goals presented in Section~\ref{subsec:system_goal}. Our security arguments build on that of threshold secret sharing scheme~\cite{shamir_secret_share}, 
Raft consensus protocol~\cite{raft}, and  Intel SGX enclave execution and its guarantees~\cite{sgx}.

We first show that the access committee $\mathtt{UAComm}$ guarantees rigid access eligibility. The eligibility of a 
function $\cal{F}$ is asserted based on the publicly verifiable certificate $\texttt{Cert}_{\mathcal{F}}$. This certificate proves the 
correct instantiation of $\cal{F}$. Given the security guarantee of Intel SGX enclave execution and its remote attestation mechanisms~
\cite{sgx_remote_attest}, nodes in $\mathtt{UAComm}$ can independently examine $\cal{F}$'s eligibility. The $
\textproc{ProcessRequest}$ protocol follows Raft~\cite{raft} to ensure that these nodes reach a consensus on such an eligibility. It 
also ties this consensus to the reconstruction of the required encryption key (i.e., acknowledgements sent by the followers contain 
their key shares). This ensures only functions that are collectively deemed as eligible may receive the encryption keys and access to the requested data capsule. 

Next, we discuss how $\textproc{ProcessRequest}$ ensures unavailability of data capsules after their expiration. 
Recall that the expiring condition of a data capsule are determined based on the access log, and that nodes 
in $\mathtt{UAComm}$ discard key shares assigned to them as soon as there is an update to the access log that leads to expiration of 
the data capsule in question (i.e., step (6) and (7) of $\textproc{ProcessRequest}$ ). 
$\textproc{ProcessRequest}$ follows Raft~\cite{raft} to ensure that the replicated access logs maintained by nodes in $
\mathtt{UAComm}$ converge, and that the key shares are collectively discarded upon expiry. Disposing the encryption key effectively render the corresponding data capsule unavailable thereafter.

It is worth noting that assuring unavailability of data capsules after their expiration requires those functions that retrieved the encryption key before the capsules' expiry do not misuse the keys or the data. In particular, they must not leak the protected data or its encryption key to any unintended party, and do not persist the data on their storage. This assurance is covered under the eligibility of the requesting function, which can be attested using SGX's attestation mechanisms~\cite{sgx_remote_attest}.

From our description of $\textproc{ProcessRequest}$, it should be clear that \codename\ attains autonomy. All access control 
operations and the expiration of the data capsule are handled in a consistent manner by independent and mutually distrusting nodes in the committee, without any explicit 
action from the data owner. In addition, so long as the adversary cannot violate our threat model (e.g., it cannot compromise for than $
\lfloor \frac{n-1}{2} \rfloor$ nodes in $\mathtt{UAComm}$, nor corrupt the enclave execution mechanisms of Intel SGX), $
\mathtt{UAComm}$ safeguards both the confidentiality of encryption keys and the legitimacy of data accesses. In other words,  
\codename\ does not expose any new attack vector on the protected data in comparison with the data owners handling the access control on their own.

\section{Evaluation}
\label{sec:eval}

\definecolor{grad0}{RGB}{235, 130, 128} 
\definecolor{grad1}{RGB}{240, 185, 147} 
\definecolor{grad2}{RGB}{245, 226, 150} 
\definecolor{grad3}{RGB}{225, 235, 185} 
\definecolor{grad4}{RGB}{220, 245, 215} 

\begin{table} \centering
\caption{Latency (ms) between different regions on GCP.}
\label{tab:GCP_latency}
\resizebox{0.45\textwidth}{!} {
\begin{tabular}{|l|r|r|r|r|}
\hline
\textbf{Zone} & \textbf{us-west1} & \textbf{us-west2} & \textbf{us-east1} & \textbf{us-east4}\\
\hline\hline
\textbf{us-west1} & 0.0 & \cellcolor{grad3} 24.7 & \cellcolor{grad1} 66.7 & \cellcolor{grad2} 59.0 \\
\hline
\textbf{us-west2} & \cellcolor{grad3} 24.7 & 0.0 & \cellcolor{grad1} 62.9 & \cellcolor{grad1} 60.5 \\
\hline
\textbf{us-east1} & \cellcolor{grad1} 66.7 & \cellcolor{grad1} 62.9 & 0.0 & \cellcolor{grad4} 12.7 \\
\hline
\textbf{us-east4} & \cellcolor{grad2} 59.1 & \cellcolor{grad1} 60.4 & \cellcolor{grad4} 12.7 & 0.0 \\
\hline

\end{tabular}
}
\end{table}

This section reports our experimental study of \codename, focusing on throughput and latency of 
$\textproc{ProcessRequest}$. We conduct the experiments following two different settings. 
In the first setting, we use an in-house (local) cluster consisting of 35 servers, each equipped with 
Intel Xeon E3-1240 2.1GHz CPUs, 32GB RAM and 2TB hard drive. Each node in the access control committee 
$\mathtt{UAComm}$ runs on a separate server. The average communication latency between any two nodes is $0.13ms$.
In the second setting, we run $\mathtt{UAComm}$ on  Google Cloud Platform (GCP), wherein each nodes is 
provisioned using a separate instance with  2 vCPUs and 8GB RAM. The nodes on GCP are distributed across 
four regions, namely Los Angeles (us-west2), Oregon (us-west1), South Carolina (us-east1) and North 
Virginia (us-east4). We detail the average communication latency between nodes in these regions in 
Table~\ref{tab:GCP_latency}.

We leverage Intel SGX SDK~\cite{sgx_sdk} to implement the trusted code base. 
We recorded the latency incurred by each SGX operation on our local cluster's CPU with SGX Enabled BIOS support. Public key operations are expensive: signing and signature verification take roughly $450 \mu s$ and $844 \mu s$, respectively. Context switching and symmetric key operations take less than $5 \mu s$. 
Remote attestation, which involves access to the IAS, takes about $250 ms$ on average. 
For the experiment on GCP, since Intel SGX is not available on the platform, we need to configure the SDK to run in simulation mode~\cite{sgx_sdk}, and inject the SGX latency measured on our local cluster into the simulation. 
Unless otherwise stated, the results reported in the following are averaged over $20$ independent runs. 

Recall that $\textproc{ProcessRequest}$ builds upon the Raft consensus protocol~\cite{raft}. One critical 
parameter to be set in Raft is the timeout~\cite{raft}. The timeout configuration necessarily takes into 
consideration the communication latency between nodes in $\mathtt{UAComm}$. Our careful analysis 
indicates that given an average communication latency of $0.13ms$ on our local cluster, the timeout 
should be chosen uniformly at random from the range $[50 - 150](ms)$. On GCP with communication latency 
ranges from $24ms$ to $66ms$, we observe the optimal timeout range to be $[150 - 250](ms)$.

\definecolor{auburn}{rgb}{0.43, 0.21, 0.1}
\definecolor{britishracinggreen}{rgb}{0.0, 0.26, 0.15}
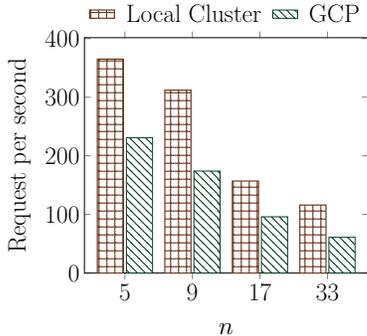
\begin{figure}[t]
\centering
\begin{tikzpicture}[thick, scale = 0.55]
\begin{axis}[
	ticklabel style = {font=\LARGE},
    ybar,
    enlarge x limits=0.2,
    legend style={at={(.0,1.1)},anchor=west, draw=none,legend columns=2, font = \LARGE, column sep=0.2cm},
    bar width=18pt,  
    area legend,
    ylabel={Request per second},   
    xlabel={$n$},   
    y label style={font = \LARGE, at={(-0.05,0.5)}},   
    x label style={font = \LARGE, at={(0.5, -0.1)}},      
    xticklabels={5, 9, 17, 33},
    xtick = data,
    ymin = 0,
    ]

\addplot[color = auburn,thick, pattern color = auburn,pattern=grid] coordinates {
(1,365) 
(2,312) 
(3,157) 
(4,116) 
};

\addplot [color = britishracinggreen, pattern color = britishracinggreen, pattern = north west lines] coordinates { 
(1,231) 
(2,174) 
(3,96) 
(4,61) 

};

\legend{Local Cluster, GCP}
\end{axis}
\end{tikzpicture}
\caption{Throughput of $\textproc{ProcessRequest}$ with respect to different access control committee size.}
\label{fig:proces_request_tp}
\end{figure}

Figure \ref{fig:proces_request_tp} reports the throughput of $\mathtt{UAComm}$ (i.e., the number of 
requests it can process per unit of time). 
We observe the correlation between the committee size and its 
throughput. More specifically, as the committee grows, the throughput drops. 
This is because larger committee leads to higher 
communication overhead of the underlying consensus protocol and higher cost of
reconstructing the encryption keys. 
We also observe that throughput recorded on GCP is lower 
than that on our local cluster. We attribute this to the difference in timeout configuration
which is influenced by the communication latency of each setting.

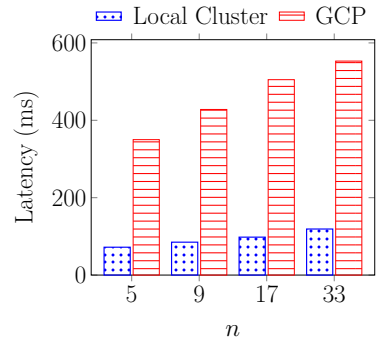
\begin{figure}[t]
\centering
\begin{tikzpicture}[thick, scale = 0.55]
\begin{axis}[
	ticklabel style = {font=\LARGE},
    ybar,
    enlarge x limits=0.2,
    legend style={at={(.0,1.1)},anchor=west, draw=none,legend columns=2, font = \LARGE, column sep=0.2cm},
    bar width=18pt,  
    area legend,
    ylabel={Latency (ms)},   
    xlabel={$n$},   
    y label style={font = \LARGE, at={(-0.05,0.5)}},   
    x label style={font = \LARGE, at={(0.5, -0.1)}},      
    xticklabels={5, 9, 17, 33},
    xtick = data,
    ymin = 0,
    ]

\addplot[color = blue,thick, pattern color = blue,pattern=dots] coordinates {
(1,72) 
(2,85) 
(3,98) 
(4,119) 
};

\addplot [color = red, pattern color = red, pattern = horizontal lines] coordinates { 
(1,350) 
(2,428) 
(3,505) 
(4,553) 

};

\legend{Local Cluster, GCP}
\end{axis}
\end{tikzpicture}
\caption{$\textproc{ProcessRequest}$  latency with respect to different access control committee size.}
\label{fig:proces_request_lt}
\end{figure}

Figure \ref{fig:proces_request_lt} depicts latency incurred by $\textproc{ProcessRequest}$ with respect 
to different access control committee size (denoted by $n$). Clearly, larger committee size leads to 
higher latency. In particular, on our local cluster, $\textproc{ProcessRequest}$ latency is at $72ms$ 
when $n=5$, which increases to $119ms$ when $n = 33$. This is due to the overhead of the underlying 
consensus protocol. Moreover, the latency observed on GCP is much higher than that observed on our local 
cluster. This is in keeping with the observation we make on the throughput difference in 
Figure~\ref{fig:proces_request_tp}, wherein the main factor contributing to the gap is the difference in  
communication latency of the two settings.

\section{Related work}
\label{sec:related_work}
The problem of self-expiring/self-destructing data capsule has been explored in the literature~\cite{fullpp, vanish}. In these works, the access policies and expiring conditions are mainly based on timing control, whereas \codename\ allows data owners to define generic conditions.
Xiong et al. proposed FullPP~\cite{fullpp}, combining timed-release encryption~\cite{cathalo2005_TRE} and 
distributed hash table (DHT) to build a full life-cycle privacy protection scheme for sensitive data. In particular, the sensitive data is first encrypted into a data-ciphertext. The key that can be used to decrypt the said ciphertext is then further encrypted into a key-ciphertex which is combined with data-ciphertext to create ciphertext shares that are distributed to the DHT network. This protocol ensures that both the key and the ciphertext are destructed after expiration time. Geambasu et al. also leverage DHT to build a self-destructing data capsule scheme called Vanish~\cite{vanish}. Unlike FullPP, Vanish only encrypts the sensitive data, splits the encryption keys into shares and distribute them to the DHT, without further encrypting the encryption key themselves or splitting the ciphertext into shares.

Standing in contrast to self-expiring data capsule is self-emerging data storage and/or time release 
encryption. In a nutshell, a sensitive data is encrypted and thus remains unreadable until a predefined 
time in the future. Ning et al.~\cite{ning2018keeping} proposed a construction that leverages threshold 
secret sharing and smart contracts. More specifically, the secret is divided into shares and entrusted to a group of incentivized participants, with rewards for punctual release of the shares contractualized and enforced by the autonomous smart contract. Li et al. developed a self-emerging data storage wherein keys of encrypted data are given to nodes in a DHT, and revealed only after a predetermined time by routing the shares in a deterministic way \cite{self_emerging}. Dragchute, on the other hand, employs time-consuming computations to incorporating timing control into the protocol \cite{dragchute}.

\section{Conclusion}
\label{sec:conclusion}
We have presented \codename, a framework for self-expiring data capsule that supports rich and generic 
access control policies and expiration conditions. The pivotal contribution underlying our proposal
is a unique integration of hardware primitives (i.e., Intel SGX), state machine replication and threshold
secret sharing in the design of \codename. 
We conduct empirical experiments on a realistic deployment setting, showing that \codename\ is able to handle access requests at scale with sub-second latency.


\bibliographystyle{abbrv}
\bibliography{paper}

\begin{thebibliography}{10}

\bibitem{self_destruct_chip}
The disintegrating computer chip.
\newblock \url{https://www.xerox.com/en-us/insights/new-computer-chip}.

\bibitem{fb_scandal}
Facebook cambridge analytica data scandal.
\newblock
  \url{https://en.wikipedia.org/wiki/Faceboo-Cambridge_Analytica_data_scandal}.

\bibitem{sgx_sdk}
Intel {SGX SDK} for {Linux}.
\newblock \url{https://github.com/01org/linux-sgx}.

\bibitem{intel_skylake}
Intel skylake processors.
\newblock \url{https://ark.intel.com/products/codename/37572/Skylake}.

\bibitem{ias}
Intel software guard extensions: Intel attestation service api.
\newblock
  \url{https://software.intel.com/sites/default/files/managed/7e/3b/ias-api-spec.pdf}.

\bibitem{leveldb}
{LevelDB}.
\newblock \url{http://leveldb.org//}.

\bibitem{panama}
Panama papers.
\newblock \url{https://www.icij.org/investigations/panama-papers/}.

\bibitem{rocksdb}
{RocksDB}.
\newblock \url{https://rocksdb.org/}.

\bibitem{SingHealth_leak}
Singhealth cyber attack: How it unfolded.
\newblock
  \url{https://graphics.straitstimes.com/STI/STIMEDIA/Interactives/2018/07/sg-cyber-breach/index.html}.

\bibitem{mscoco}
{The Coco Framework}.
\newblock \url{http://aka.ms/cocopaper}.

\bibitem{trustzone}
Tiago Alves.
\newblock Trustzone: Integrated hardware and software security.
\newblock {\em ARM}, 2004.

\bibitem{sgx_remote_attest}
Ittai Anati, Shay Gueron, Simon Johnson, and Vincent Scarlata.
\newblock Innovative technology for cpu based attestation and sealing.
\newblock In {\em Proceedings of the 2nd international workshop on hardware and
  architectural support for security and privacy}, volume~13. ACM New York, NY,
  USA, 2013.

\bibitem{concurrency_control_book}
Philip~A Bernstein, Vassos Hadzilacos, and Nathan Goodman.
\newblock {\em Concurrency control and recovery in database systems}.
\newblock Addison-wesley New York, 1987.

\bibitem{rollback_detection}
Marcus Brandenburger, Christian Cachin, Matthias Lorenz, and R{\"u}diger
  Kapitza.
\newblock Rollback and forking detection for trusted execution environments
  using lightweight collective memory.
\newblock In {\em DSN}, 2017.

\bibitem{pbft}
Miguel Castro, Barbara Liskov, et~al.
\newblock Practical byzantine fault tolerance.
\newblock In {\em OSDI}, volume~99, pages 173--186, 1999.

\bibitem{cathalo2005_TRE}
Julien Cathalo, Beno{\^\i}t Libert, and Jean-Jacques Quisquater.
\newblock Efficient and non-interactive timed-release encryption.
\newblock In {\em International Conference on Information and Communications
  Security}, pages 291--303. Springer, 2005.

\bibitem{sanctum}
Victor Costan, Ilia Lebedev, and Srinivas Devadas.
\newblock Sanctum: Minimal hardware extensions for strong software isolation.
\newblock In {\em 25th $\{$USENIX$\}$ Security Symposium ($\{$USENIX$\}$
  Security 16)}, 2016.

\bibitem{AMES}
Hung Dang and Ee-Chien Chang.
\newblock Autonomous membership service for enclave applications.
\newblock {\em arXiv preprint arXiv:1905.06460}, 2019.

\bibitem{STC}
Hung Dang, Tien Tuan~Anh Dinh, Ee-Chien Chang, and Beng~Chin Ooi.
\newblock Privacy-preserving computation with trusted computing via
  scramble-then-compute.
\newblock {\em Proceedings on Privacy Enhancing Technologies}, 2017(3):21--38,
  2017.

\bibitem{sigmod_sharding}
Hung Dang, Tien Tuan~Anh Dinh, Dumitrel Loghin, Ee-Chien Chang, Qian Lin, and
  Beng~Chin Ooi.
\newblock Towards scaling blockchain systems via sharding.
\newblock In {\em Proceedings of the 2019 International Conference on
  Management of Data}, 2019.

\bibitem{differential_privacy}
Cynthia Dwork, Aaron Roth, et~al.
\newblock The algorithmic foundations of differential privacy.
\newblock {\em Foundations and Trends{\textregistered} in Theoretical Computer
  Science}, 2014.

\bibitem{FLP_impossibility}
Michael~J Fischer, Nancy~A Lynch, and Michael~S Paterson.
\newblock Impossibility of distributed consensus with one faulty process.
\newblock Technical report, Massachusetts Inst of Tech Cambridge lab for
  Computer Science, 1982.

\bibitem{fraser2000hardening}
Timothy Fraser, Lee Badger, and Mark Feldman.
\newblock Hardening cots software with generic software wrappers.
\newblock In {\em Proceedings DARPA Information Survivability Conference and
  Exposition. DISCEX'00}. IEEE.

\bibitem{vanish}
Roxana Geambasu, Tadayoshi Kohno, Amit~A. Levy, and Henry~M. Levy.
\newblock Vanish: Increasing data privacy with self-destructing data.
\newblock In {\em USENIX Security Symposium}, 2009.

\bibitem{epid}
Simon Johnson, Vinnie Scarlata, Carlos Rozas, Ernie Brickell, and Frank Mckeen.
\newblock Intel software guard extensions: Epid provisioning and attestation
  services.
\newblock {\em White Paper}, 1:1--10, 2016.

\bibitem{DHT}
M~Frans Kaashoek and David~R Karger.
\newblock Koorde: A simple degree-optimal distributed hash table.
\newblock In {\em International Workshop on Peer-to-Peer Systems}, pages
  98--107. Springer, 2003.

\bibitem{crypto_intro}
Jonathan Katz and Yehuda Lindell.
\newblock {\em Introduction to modern cryptography}.
\newblock Chapman and Hall/CRC, 2014.

\bibitem{keystone}
Dayeol Lee, David Kohlbrenner, Shweta Shinde, Dawn Song, and Krste Asanovic.
\newblock Keystone: An open framework for architecting tees, 2019.

\bibitem{self_emerging}
C.~{Li} and B.~{Palanisamy}.
\newblock Timed-release of self-emerging data using distributed hash tables.
\newblock In {\em 2017 IEEE 37th International Conference on Distributed
  Computing Systems (ICDCS)}, pages 2344--2351, June 2017.

\bibitem{churp}
Sai Krishna~Deepak Maram, Fan Zhang, Lun Wang, Andrew Low, Yupeng Zhang, Ari
  Juels, and Dawn Song.
\newblock Churp: Dynamic-committee proactive secret sharing.
\newblock {\em IACR Cryptology ePrint Archive}, 2019.

\bibitem{rote}
Sinisa Matetic, Mansoor Ahmed, Kari Kostiainen, Aritra Dhar, David Sommer,
  Arthur Gervais, Ari Juels, and Srdjan Capkun.
\newblock Rote: Rollback protection for trusted execution.
\newblock {\em IACR Cryptology ePrint Archive}, 2017.

\bibitem{sgx}
Frank McKeen, Ilya Alexandrovich, Alex Berenzon, Carlos~V Rozas, Hisham Shafi,
  Vedvyas Shanbhogue, and Uday~R Savagaonkar.
\newblock Innovative instructions and software model for isolated execution.
\newblock {\em HASP@ ISCA}, 10, 2013.

\bibitem{btc_origin}
Satoshi Nakamoto.
\newblock Bitcoin: A peer-to-peer electronic cash system, 2008.

\bibitem{ning2018keeping}
Jianting Ning, Hung Dang, Ruomu Hou, and Ee-Chien Chang.
\newblock Keeping time-release secrets through smart contracts.
\newblock {\em IACR Cryptology ePrint Archive}, 2018:1166, 2018.

\bibitem{raft}
Diego Ongaro and John~K Ousterhout.
\newblock In search of an understandable consensus algorithm.
\newblock In {\em USENIX Annual Technical Conference}, pages 305--319, 2014.

\bibitem{RSM}
Fred~B Schneider.
\newblock Implementing fault-tolerant services using the state machine
  approach: A tutorial.
\newblock {\em ACM Computing Surveys (CSUR)}, 22, 1990.

\bibitem{shamir_secret_share}
Adi Shamir.
\newblock How to share a secret.
\newblock {\em Communications of the ACM}, 1979.

\bibitem{nonblocking_commit}
Dale Skeen.
\newblock Nonblocking commit protocols.
\newblock In {\em Proceedings of the 1981 ACM SIGMOD international conference
  on Management of data}, 1981.

\bibitem{stytz2004considering}
Martin~R Stytz.
\newblock Considering defense in depth for software applications.
\newblock {\em IEEE Security \& Privacy}, 2004.

\bibitem{enclave_formalization}
Pramod Subramanyan, Rohit Sinha, Ilia Lebedev, Srinivas Devadas, and Sanjit~A
  Seshia.
\newblock A formal foundation for secure remote execution of enclaves.
\newblock In {\em Proceedings of the 2017 ACM SIGSAC Conference on Computer and
  Communications Security}, pages 2435--2450. ACM, 2017.

\bibitem{dragchute}
Luis Vargas, Gyan Hazarika, Rachel Culpepper, Kevin~R.B. Butler, Thomas
  Shrimpton, Doug Szajda, and Patrick Traynor.
\newblock Mitigating risk while complying with data retention laws.
\newblock In {\em Proceedings of the 2018 ACM SIGSAC Conference on Computer and
  Communications Security}, CCS '18, pages 2011--2027, New York, NY, USA, 2018.
  ACM.

\bibitem{fullpp}
Jinbo Xiong, Fenghua Li, Jianfeng Ma, Ximeng Liu, Zhiqiang Yao, and Patrick~S.
  Chen.
\newblock A full lifecycle privacy protection scheme for sensitive data in
  cloud computing.
\newblock {\em Peer-to-Peer Networking and Applications}, 8(6):1025--1037, Nov
  2015.

\bibitem{controlled_channel}
Yuanzhong Xu, Weidong Cui, and Marcus Peinado.
\newblock Controlled-channel attacks: Deterministic side channels for untrusted
  operating systems.
\newblock In {\em Security and Privacy (SP), 2015 IEEE Symposium on}, pages
  640--656. IEEE, 2015.

\bibitem{safevanish}
Lingfang Zeng, Zhan Shi, Shengjie Xu, and Dan Feng.
\newblock Safevanish: An improved data self-destruction for protecting data
  privacy.
\newblock In {\em 2010 IEEE Second International Conference on Cloud Computing
  Technology and Science}, 2010.

\end{thebibliography}

\end{document}